\pdfoutput=1

\documentclass[11pt]{article}

\usepackage[preprint]{acl}

\usepackage{times}
\usepackage{booktabs}      
\usepackage{multirow}
\usepackage{float}

\usepackage[T1]{fontenc}

\usepackage{hyperref}

\usepackage[utf8]{inputenc}

\usepackage{microtype}

\usepackage{inconsolata}

\usepackage{graphicx}

\title{RT-VC: Real-Time Zero-Shot Voice Conversion\\ with Speech Articulatory Coding}

\author{Yisi Liu, Chenyang Wang, Hanjo Kim, Raniya Khan, Gopala Anumanchipalli \\
        UC Berkeley \\
        \href{https://berkeley-speech-group.github.io/RT-VC/}{\texttt{https://berkeley-speech-group.github.io/RT-VC/}}}

\begin{document}
\maketitle
\begin{abstract}
Voice conversion has emerged as a pivotal technology in numerous applications ranging from assistive communication to entertainment. In this paper, we present RT-VC, a zero-shot real-time voice conversion system that delivers ultra-low latency and high-quality performance. Our approach leverages an articulatory feature space to naturally disentangle content and speaker characteristics, facilitating more robust and interpretable voice transformations. Additionally, the integration of differentiable digital signal processing (DDSP) enables efficient vocoding directly from articulatory features, significantly reducing conversion latency. Experimental evaluations demonstrate that, while maintaining synthesis quality comparable to the current state-of-the-art (SOTA) method, RT-VC achieves a CPU latency of 61.4 ms, representing a 13.3\% reduction in latency. 

\end{abstract}

\section{Introduction}

Voice conversion (VC) modifies speech to match the timbre of a target speaker while preserving content information. A central challenge in VC is the effective disentanglement of speaker identity from the underlying content. This separation is critical to enable the transformation of voice characteristics while maintaining the linguistic and paralinguistic information, including emotion and accent.

There are three principal strategies to achieve disentanglement between speaker and content representations in voice conversion. First, autoencoder‐based approaches employ encoder–decoder architectures (often variational) and incorporate carefully designed bottlenecks or specialized modules to isolate speaker identity from linguistic content \citep{autovc, triple_bottleneck, naturalspeech3, d-dsvae, instance_norm}. Second, GAN‐based methods leverage generative adversarial networks and domain-mapping losses (e.g., cycle-consistency) to ensure that the converted speech retains the source content while convincingly mimicking the target speaker’s characteristics \citep{cyclegan-vc, cyclegan-vc2, stargan-vc, stargan-vc2, peter-vc}. Third, methods leveraging pretrained models for representation learning extract speaker-independent content representations from external systems, such as automatic speech recognition (ASR) \citep{ppg-vc, hifivc, cosyvoice, cosyvoice2}, text-to-speech (TTS) \citep{cotatron}, or self-supervised learning frameworks \citep{softvc, streamvc, nansy, contentvec, freevc}.

While these methods achieve impressive performance, they often require meticulous architectural design and careful tuning of loss functions. Moreover, they typically operate as black-box models, relying on abstract latent spaces that lack interpretability and universality. To address these limitations and achieve a more natural, straightforward, and grounded disentanglement between speaker and content representations, we adopt the Speech Articulatory Coding (SPARC) framework \citep{sparc}. In SPARC, content information is represented as vocal tract kinematics within a normalized, speaker-agnostic space, while speaker-specific characteristics are captured separately via a dedicated speaker encoder. This approach yields a naturally disentangled and interpretable representation that supports accent-preserving, zero-shot voice conversion. However, the transformation between speech and the articulatory feature space is computationally intensive, making SPARC less suitable for real-time applications.

In this paper, we present RT-VC, a zero-shot real-time voice conversion system that combines SPARC with efficient streaming architecture. In order to accelerate the SPARC encoding process (speech to articulatory features), we train a causal source extractor and a causal acoustic-to-articulatory inversion model using the labels from SPARC encoding. For SPARC decoding (articulatory features to speech), we utilize the differentiable digital signal processing (DDSP) vocoder from \citep{slt2024}, which is known for fast inference and high quality. Our experimental results show that RT-VC achieves intelligibility and speaker similarity comparable to the current SOTA real-time zero-shot voice conversion system, StreamVC \citep{streamvc}. In addition, RT-VC achieves an end-to-end CPU latency of 61.4ms, which is 13.3\% faster than StreamVC.

\section{Related Work}
\begin{figure*}[t]
    \centering
    \includegraphics[width=\linewidth]{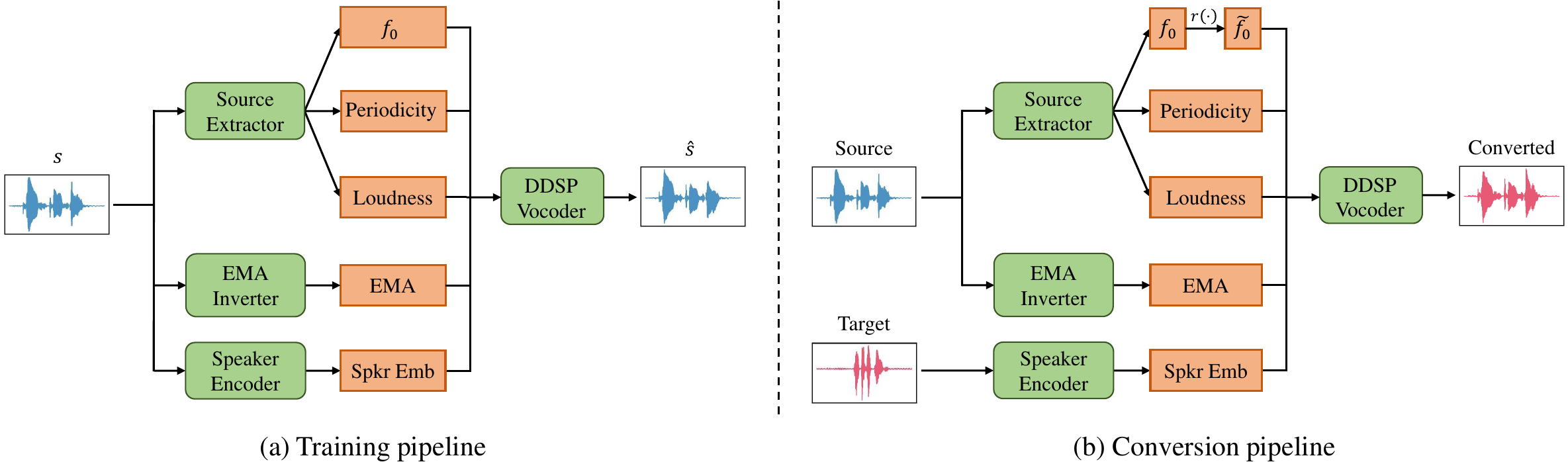}
    \caption{Training and conversion pipeline of RT-VC. $s$ denotes input speech, $\hat{s}$ denotes reconstructed speech, $r(\cdot)$ denotes the pitch rescaling operation in (\ref{rescaling}).}
    \label{system}
\end{figure*}

\subsection{Zero-Shot Voice Conversion}
Zero-shot voice conversion refers to converting speech from a source speaker to the voice of a new, previously unseen target speaker without requiring any parallel or fine-tuning data for that speaker during training. Achieving this requires a precise disentanglement of speaker characteristics from the linguistic content.

One of the earliest approaches in this domain is AUTOVC \citep{autovc}, which employs an autoencoder architecture with a carefully designed bottleneck to preserve content information while stripping away speaker-specific features. This bottleneck concept is also demonstrated in NaturalSpeech 3 \citep{naturalspeech3}, where separate bottlenecks for prosody, content, and acoustic details are constructed to remove unnecessary information and facilitate disentanglement.

In contrast, the StarGAN-VC family \citep{stargan-vc, stargan-vc2} formulates voice conversion as a domain translation problem between speaker domains. These methods utilize a combination of GAN loss and content preservation loss to guide the model to modify only speaker-related features.

Recent approaches utilize pretrained models for obtaining content representations. For instance, HiFi-VC \citep{hifivc} uses bottleneck features from a pretrained ASR system as the content representation, while the CosyVoice family \citep{cosyvoice, cosyvoice2} further quantizes the ASR bottleneck features to enhance disentanglement. Cotatron \citep{cotatron} utilizes a pretrained autoregressive TTS model to provide text-speech alignment and employs the aligned phoneme features as content representations. Additionally, SoftVC \citep{softvc} and StreamVC \citep{streamvc} leverage the self-supervised learning model HuBERT \citep{hubert} to derive discrete labels via k-means clustering; a content encoder is then trained to predict these labels, with the resulting continuous features serving as the content representation. NANSY \citep{nansy} employs information perturbation techniques to isolate linguistic information from wav2vec 2.0 \cite{wav2vec2}, and ContentVec \citep{contentvec} applies the same techniques to HuBERT.

\subsection{Acoustic-to-Articulatory Inversion}
Acoustic-to-articulatory inversion (AAI) aims to predict vocal tract kinematics from raw speech, with these kinematics typically measured via electromagnetic articulography (EMA). EMA captures distinct patterns of articulator movements that naturally encode linguistic content \citep{ppg-vc, sparc}. However, the scalability of EMA is limited by the high costs of data collection and its inherent entanglement with speaker-specific anatomical features. Recent AAI models \citep{peter_aai, vt_lab, tv_aai, source_aai} have been proposed to alleviate the collection burden, but they do not fully resolve the issue of speaker entanglement. To address this, \citep{universal-aai, sparc} argue that the differences between individual speakers’ articulatory systems can be approximated by a single linear affine transformation, and propose the use of a universal articulatory space derived from a single speaker as a common template for all speakers. These insights provide the foundation for developing voice conversion systems that leverage articulatory features to disentangle linguistic content from speaker characteristics.

\subsection{Articulatory Synthesis}
Articulatory synthesis, the inverse task of acoustic-to-articulatory inversion (AAI), involves generating speech from articulatory features like EMA. Recent deep learning approaches in this domain have predominantly employed GAN-based vocoders like HiFi-GAN \citep{hifigan} to synthesize speech either from intermediate spectrograms \citep{EMA2S, style} or directly from articulatory inputs \citep{Peter-ATS, sparc}. A recent study \citep{slt2024} utilizes differentiable digital signal processing (DDSP) to achieve fast inference, high quality and improved parameter efficiency. In our work, we adopt the DDSP vocoder from \citep{slt2024} to enable real-time voice conversion.

\section{Method}
In this section, we first present an overview of the complete system during both training and inference (Section \ref{overview}). Next, we describe the architecture and training strategies for each module of the system (Sections \ref{source_extractor} through \ref{vocoder}). Finally, we outline the streaming strategy for real-time voice conversion (Section \ref{streaming}).

\subsection{System Overview}\label{overview}
Building on the framework presented in \citep{sparc}, our proposed system comprises four primary components: a source extractor, an EMA inverter, a speaker encoder, and a DDSP vocoder. With the exception of the offline speaker encoder, all components are designed to be streamable. 

An overview of the complete system architecture is provided in Figure \ref{system}. During training, the input speech signal is decomposed into an articulatory feature space comprising pitch, periodicity, loudness, EMA, and speaker embedding. The DDSP vocoder then reconstructs the speech signal from these features. Notably, the source extractor and EMA inverter are initially trained independently of the whole system. Subsequently, the speaker encoder and DDSP vocoder are jointly optimized using the outputs of the two pretrained modules. During conversion, the speaker embedding is extracted from the target speaker's utterance, and the source pitch is adjusted to match the target speaker's range by scaling it with the ratio of the target speaker's median pitch ($m_{tgt}$) to the source speaker's median pitch ($m_{src}$):
\begin{equation}
    \tilde{f_0} = r(f_0) = f_0\cdot\frac{m_{tgt}}{m_{src}}
    \label{rescaling}
\end{equation}

\subsection{Source Extractor}\label{source_extractor}
The source extractor is designed to isolate laryngeal source information from the input speech. Specifically, it extracts source features including pitch (indicative of the vocal fold vibration frequency), periodicity (reflecting the presence or absence of vocal fold oscillation), and loudness (representing the energy of the airflow through the larynx).

We reformulate the pitch tracking problem as a frequency bin classification task, following the approach outlined in \citep{crepe, rmvpe}. In our method, the source extractor accepts a mel spectrogram as input and generates an encoding using a series of causal convolution blocks following the SoundStream encoder architecture \citep{soundstream}. This encoding is then processed by three distinct linear output layers: a pitch head that transforms the encoding into a probability distribution over all potential frequency bins for each time frame, a periodicity head that determines whether each input frame is voiced or unvoiced, and a loudness head that predicts the frame-level energy. To get the final pitch prediction, we use the local weighted average of frequencies closest to the frequency bin with the highest probability, as described in \citep{rmvpe}. Although a simple digital signal processing method such as a moving average could be used to estimate loudness, we have found that such an approach is highly sensitive to noise. Therefore, we utilize a dedicated loudness head to produce a clean loudness estimate even under noisy conditions, thereby enhancing the overall noise robustness of the system.
To obtain ground truth labels for pitch and periodicity, we employ CREPE \citep{crepe} to generate the pitch values and RMVPE \citep{rmvpe} to derive binary voiced flags. Loudness labels are computed by averaging the clean input spectrogram along the frequency axis. Pitch, voiced flags and loudness are all sampled at 200Hz. We follow the cross entropy loss introduced in \citep{rmvpe} to train our pitch and periodicity heads. For loudness head, a simple L1 loss between the prediction and the ground truth is applied. To enhance the noise robustness of our source extractor, we add noise augmentation using the \texttt{audiomentation} package\footnote{\href{https://github.com/iver56/audiomentations}{https://github.com/iver56/audiomentations}}. Specifically, we utilize the \texttt{AddColorNoise} module to introduce noise with varied spectral characteristics and the \texttt{RoomSimulator} module to apply different room impulse responses.

\subsection{EMA Inverter}
We train a real-time EMA inverter based on the SoundStream encoder architecture \citep{soundstream}. It takes MFCC as input, and processes the input features through 11 dilated causal convolution layers followed by an MLP to get the predicted EMA output. 

We also add augmentation during EMA inverter training. Prior to applying noise augmentation, we adopt the information perturbation technique proposed in \citep{nansy}, which sequentially applies a random parametric equalizer, pitch randomization, and formant shifting. Since these operations preserve content-level information, they encourage the EMA inverter to focus primarily on content features, thereby promoting improved disentanglement from speaker-specific characteristics. 

To get the EMA ground truth, we generate pseudo EMA labels using the acoustic-to-articulatory inversion model from \citep{sparc, universal-aai, ssl-aai}. We linearly interpolate these pseudo EMA from 50Hz to 200Hz. The EMA inverter is trained to minimize the L1 loss between the predicted EMA and the pseudo EMA labels.

\subsection{Speaker Encoder}
Similar to \citep{sparc}, our speaker encoder contains a frozen CNN feature extractor of WavLM \citep{wavlm} and a trainable dilated convolution network. The output encoding will be aggregated into a 128-dimensional speaker embedding using the periodicity output from the pretrained source extractor as the weight. The speaker encoder is trained together with the vocoder. 

\subsection{DDSP Vocoder}\label{vocoder}
\begin{figure}[t]
    \centering
    \includegraphics[width=\linewidth]{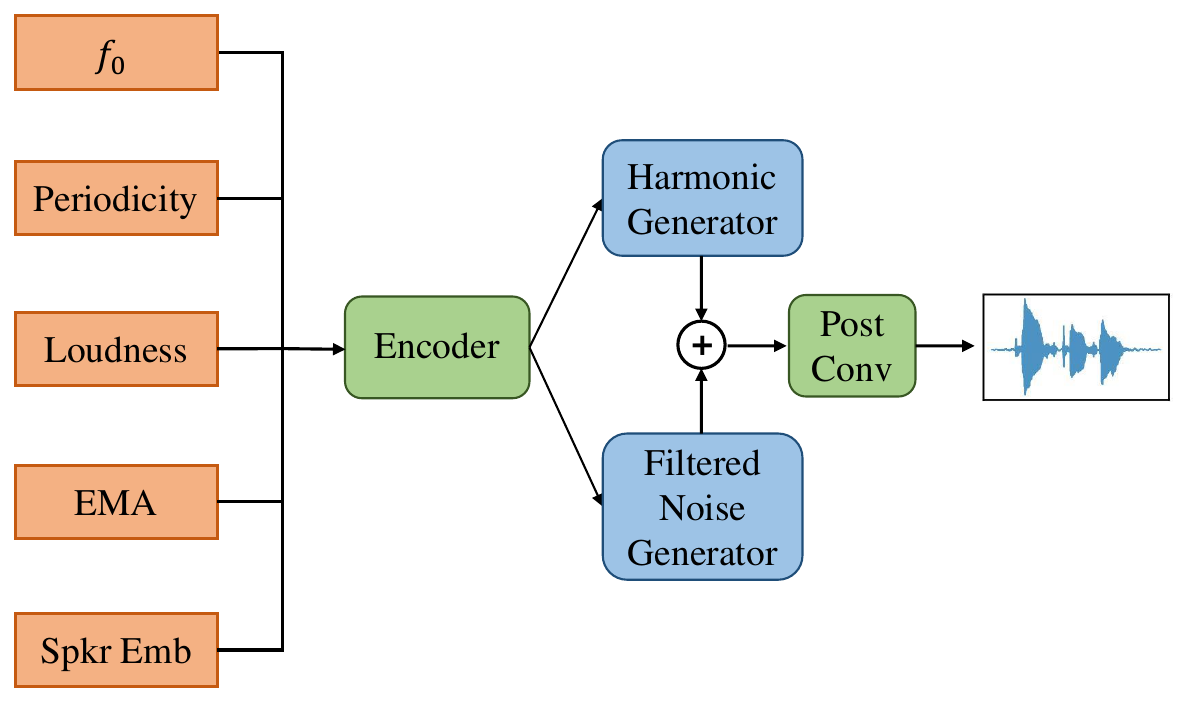}
    \caption{DDSP vocoder architecture. }
    \label{vocoder_fig}
\end{figure}

We adopt the DDSP harmonic-plus-noise vocoder from \citep{slt2024} to enable fast inference. The model architecture is shown in Figure \ref{vocoder_fig}. The encoder accepts the previously described articulatory features as input and separately predicts control signals for harmonic generator and filtered noise generator to generate periodic (harmonic) and aperiodic (noise) components. These components are summed and then filtered through a post convolution layer to produce the final speech output. To condition the vocoder on speaker-specific characteristics, we integrate a FiLM layer \citep{film} that processes the speaker embedding and produces scaling and shifting parameters to modulate the intermediate encoding. To make the vocoder streamable, we use the SoundStream encoder architecture \cite{soundstream} with 11 dilated causal convolution layers. The post convolution layer is also made causal. We train the model using the loss functions described in \citep{slt2024}, namely, the multi-scale spectral loss and the multi-resolution adversarial loss.

\subsection{Real-Time Inference}\label{streaming}
For real-time inference, the input spectral features (mel spectrogram and MFCC) are calculated on the fly. The window size is chosen to be 1024 at 16kHz for all spectral features, with reflection padding to center each output frame. This translates into a lookahead of half the window size, i.e. 32ms. Since our system is causal, we only need to maintain a ring buffer to store the running past context for each module during streaming, where the length of the context is determined by the receptive field of the causal convolution network. Additionally, to facilitate pitch rescaling to the target speaker's range, a running median of the source pitch is also maintained.

The end-to-end latency $L$ is calculated as:
\begin{equation}
    L = t_{lookahead}+t_{chunksize}+t_{processing}
\end{equation}
Here $t_{lookahead}$ = 32ms (half the window size), $t_{chunksize}$ = 15ms (the input chunk size), and $t_{processing}$ = 14.4ms is the average processing time for each chunk on an Apple M3 CPU. Therefore, the end-to-end latency is 61.4ms, which is faster than the current SOTA (StreamVC, 70.8ms) by 13.3\%.

\section{System Design}
\begin{figure}[t]
  \centering
  \includegraphics[width=\linewidth]{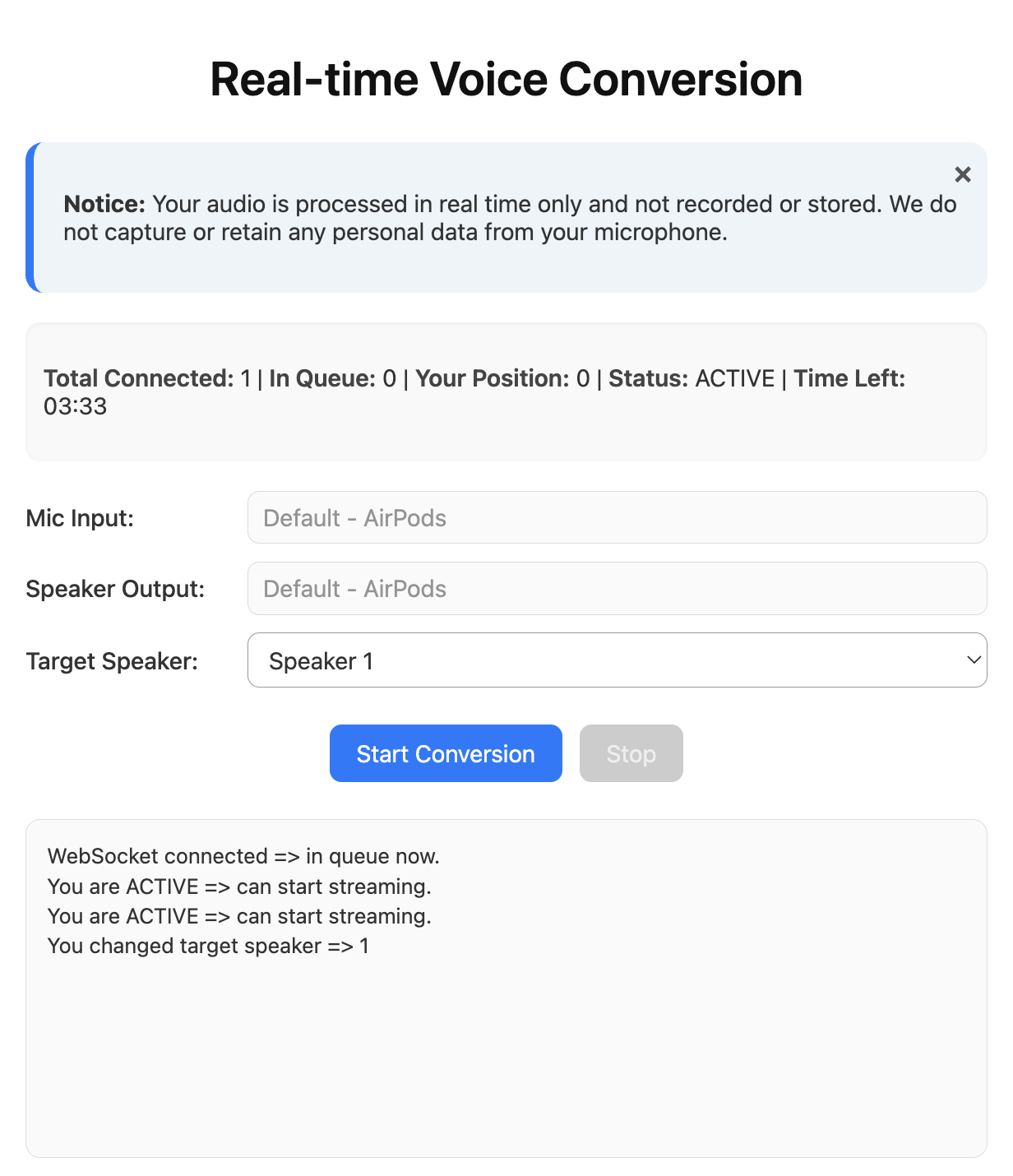}
  \caption{Screenshot of the RT-VC web demo interface.}
  \label{web_demo}
\end{figure}

A screen shot of the RT-VC web demo is shown in Figure \ref{web_demo}. This demo enables real-time voice conversion directly through the web interface, eliminating the need for any downloads. During conversion, the user speaks into the frontend, where the incoming audio is sampled at 16 kHz and segmented into 15ms chunks. These chunks are then transmitted to the backend for real-time inference (see Section \ref{streaming}), and the converted audio is returned to the frontend for playback through the designated output device.

For audio input and output, they are configured to use the system's default devices. We recommend using a high-quality microphone with echo cancellation to minimize input noise and reduce speaker feedback. If necessary, users may modify their audio device settings via the system configuration and refresh the webpage to apply the changes. 

For target speaker selection, users may choose from 10 pre-enrolled target speakers drawn from the VCTK dataset \citep{vctk}, with all target speakers being unseen during training. Moreover, the system allows users to dynamically switch the target speaker while speaking, with the generated voice updating instantly.

The web demo is deployed on an AWS CPU server (C7i instance type) equipped with an Intel Xeon Scalable processor. Due to CPU resource constraints, only one user can access the web demo at a time for at most 5 minutes. Additional users are queued and notified when their session begins.

\begin{table*}[t] 
    \centering
    \label{tab:metrics}
    \renewcommand{\arraystretch}{1.3} 
    \resizebox{\textwidth}{!}{ 
    \begin{tabular}{l cc cc cc c c}
        \toprule
        \multirow{2}{*}{\raggedright\textbf{Model Name}} 
        & \multicolumn{2}{c}{\textbf{Naturalness}} 
        & \multicolumn{2}{c}{\textbf{Intelligibility}} 
        & \multicolumn{2}{c}{\textbf{Speaker Similarity}} 
        & \textbf{$f_0$ Consistency}
        & \textbf{CPU} \\
        \cmidrule(lr){2-3} 
        \cmidrule(lr){4-5} 
        \cmidrule(lr){6-7} 
        \cmidrule(lr){8-8}

        & \textbf{{UTMOS} $\uparrow$}& \textbf{{MOS} $\uparrow$}
         & \textbf{{WER} $\downarrow$} & \textbf{{CER} $\downarrow$}
         & \textbf{{Resemblyzer Score} $\uparrow$} & \textbf{{SMOS} $\uparrow$}
         & \textbf{{$f_0$ PCC} $\uparrow$} 
         & \textbf{{Latency} $\downarrow$}\\
         
        \midrule
        \textbf{Source (LibriTTS)} & 4.03 $\pm$ 0.04 & 4.13 $\pm$ 0.16 & 5.06\% & 1.36\% & - & - & - & - \\
        \midrule
        \textbf{StreamVC}  & - & - & \textbf{6.22\%} & 2.17\% & \textbf{77.81\%} & - & 0.842 & 70.8ms \\
        \textbf{RT-VC} & 3.81 $\pm$ 0.02 & 3.87 $\pm$ 0.17 & 6.69\% & \textbf{2.12\%} & 76.65\% & 3.59 $\pm$ 0.19 & \textbf{0.865} & \textbf{61.4ms} \\
        \bottomrule
    \end{tabular}
    }
    \caption{Performance comparison of StreamVC and RT-VC. StreamVC values are taken directly from its publication. Values are presented with their corresponding 95\% confidence intervals where applicable.}
\end{table*}

\section{Results}

\subsection{Dataset}
Each module of the system is trained on the \texttt{train} subset of LibriTTS-R \citep{libritts_r}, which is a restored version of LibriTTS \citep{libritts}. The \texttt{train} subset contains 555 hours of speech from 2311 speakers. All samples are downsampled to 16kHz. 

For evaluation and direct comparison with the current SOTA StreamVC, we use the same test set: we extract 377 source utterances from the \texttt{test-clean} subset of LibriTTS and select 6 target speakers from VCTK \citep{vctk}. Importantly, all source and target speakers are unseen during training, thereby assessing the zero-shot voice conversion capability of the systems. 

\subsection{Metrics}
We evaluate the models along four key dimensions: naturalness, intelligibility, speaker similarity, and $f_0$ consistency. Since StreamVC is not open source, to enable a direct comparison with StreamVC, we adopt the same evaluation protocol for all metrics except for naturalness and speaker similarity, as StreamVC did not incorporate subjective evaluation for these aspects. In addition, we were unable to reproduce the naturalness results using the official DNSMOS\footnote{\href{https://github.com/microsoft/DNS-Challenge}{https://github.com/microsoft/DNS-Challenge}} repository because the upper bound of DNSMOS for clean speech appears to be around 3.3\footnote{\href{https://github.com/microsoft/DNS-Challenge/issues/189}{https://github.com/microsoft/DNS-Challenge/issues/189}}, whereas StreamVC reports a DNSMOS of 3.99 for source utterances from LibriTTS. Consequently, we use alternative, widely used metrics for naturalness evaluation.

Naturalness is measured automatically using UTMOS\footnote{\href{https://github.com/sarulab-speech/UTMOS22}{https://github.com/sarulab-speech/UTMOS22}}, which is a machine-evaluated mean opinion score (MOS), and subjectively via a 5-point MOS test on Prolific\footnote{\href{https://www.prolific.com/}{https://www.prolific.com/}}. Each model receives 200 unique ratings. Intelligibility is evaluated using word error rate (WER) and character error rate (CER), both obtained using the \texttt{HuBERT-Large} ASR model\footnote{\href{https://huggingface.co/facebook/hubert-large-ls960-ft}{https://huggingface.co/facebook/hubert-large-ls960-ft}}. Speaker similarity is measured automatically by the cosine similarity between speaker embeddings generated by \texttt{Resemblyzer}\footnote{\href{https://github.com/resemble-ai/Resemblyzer}{https://github.com/resemble-ai/Resemblyzer}}, and subjectively by similarity mean opinion score (SMOS) ratings from human raters. Lastly, $f_0$ consistency is evaluated using the Pearson correlation coefficient (PCC) between source and converted speech $f_0$ contours. 

\subsection{Conversion Quality}
Table 1 summarizes the performance of RT-VC and StreamVC. Overall, the two models exhibit comparable conversion quality. For naturalness, RT-VC achieves a UTMOS of $3.81$ and a MOS of $3.87$, both of which are greater than 3.8, which is a good indicator of high fidelity. For intelligibility, RT-VC performs similarly to StreamVC, with a slightly higher WER (+0.47\%) and a marginally lower CER (–0.05\%), and both metrics are close to those of the ground truth. This indicates that the converted speech of RT-VC is highly intelligible. Additionally, both systems demonstrate comparable speaker similarity and $f_0$ consistency, with RT-VC showing a slightly lower Resemblyzer score (–1.16\%) and a marginally higher $f_0$ consistency (+0.023), underscoring its strong zero-shot conversion capability.

\subsection{Noise Robustness}
\begin{figure}[t]
    \centering
    \includegraphics[width=\linewidth]{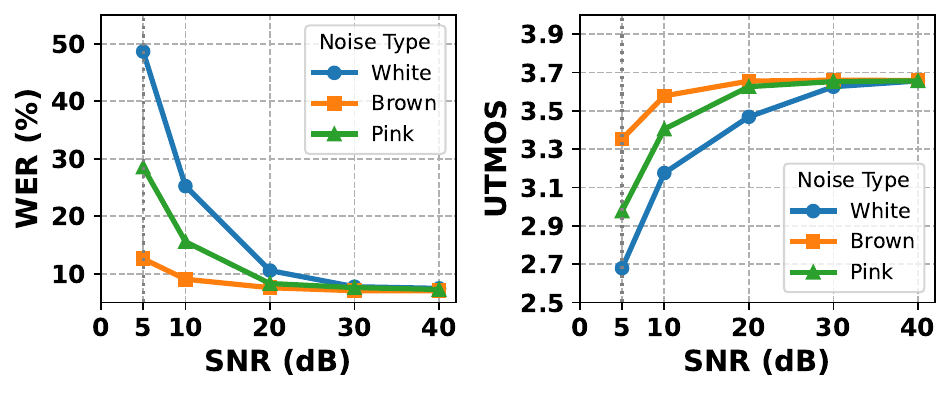}
    \caption{WER and UTMOS against input SNR for three types of additive noise: white, brown, and pink.}
    \label{noise_robustness}
\end{figure}

We assess the noise robustness of RT-VC by measuring the WER and UTMOS of the converted speech when the input source is contaminated with noise while the target remains clean. Noisy source speech is generated by adding white, pink, and brown noise at various signal-to-noise ratios (SNRs) to the original utterances. Figure \ref{noise_robustness} presents the results. Overall, RT-VC is most robust to brown noise, with minimal degradation in WER and UTMOS as the SNR decreases from 40dB to 10dB. In contrast, white noise has the greatest impact: at 20dB SNR, the WER is around 10\%, but it increases sharply to approximately 25\% at 10dB. Moreover, UTMOS drops below 3.5 when the SNR is lower than 20dB. These findings indicate that RT-VC effectively handles static noise when the input SNR is above 20dB, demonstrating strong noise robustness.

\section{Conclusion}
We introduce RT-VC, a zero-shot real-time voice conversion system that delivers low CPU latency and high conversion quality. RT-VC leverages the Speech Articulatory Coding (SPARC) framework in conjunction with a real-time DDSP vocoder, enabling natural speaker-content disentanglement with rapid conversion. Compared with the current SOTA, RT-VC achieves lower CPU latency while maintaining comparable conversion quality, and it demonstrates robustness against static background noise. Future work will explore prompt-free real-time voice conversion by incorporating offline design of target speaker characteristics, such as gender, age, emotion, and accent.

\section{Limitations}
RT-VC leverages the Speech Articulatory Coding (SPARC) framework to enable natural and grounded disentanglement between speaker and content representations. However, there are still limitations. First, relying solely on electromagnetic articulography (EMA) does not fully capture vocal tract kinematics, as it omits crucial dynamics such as nasal cavity movements and laryngeal behavior, which are vital for modeling nasal sounds and larynx-specific phenomena like vocal fry. Second, the pseudo EMA labels are generated by a self-supervised learning model (WavLM) that was pretrained exclusively on English data and probed onto an English speaker's articulation space. Although our video demonstration shows that the system can perform cross-lingual conversion, this language-specific EMA inversion restricts the model's multilingual capabilities. Third, despite training with static noise augmentation, the system remains sensitive to the quality of the input speech, and conversion performance is ultimately constrained by the recording equipment's quality.

\section{Ethical Considerations}
The ethical concerns surrounding RT-VC arise from the broader risks associated with voice conversion and generative speech models, notably the potential for impersonation and privacy violations. To mitigate these risks, RT-VC checkpoints will not be made open source, thereby limiting unrestricted access to the technology. In addition to this initial safeguard, we plan to implement further measures to prevent misuse. In particular, developing robust detection mechanisms is a priority, as these can help identify and deter unauthorized applications. Furthermore, we intend to explore the integration of watermarking or traceable metadata into the converted audio, facilitating tracking and accountability in instances of unethical use.

\section{Acknowledgement}
This work is funded in part by the IARPA ARTS program.



\bibliography{ref}

\begin{thebibliography}{42}
\providecommand{\natexlab}[1]{#1}

\bibitem[{Attia et~al.(2024)Attia, Siriwardena, and Espy-Wilson}]{tv_aai}
Ahmed~Adel Attia, Yashish~M Siriwardena, and Carol Espy-Wilson. 2024.
\newblock Improving speech inversion through self-supervised embeddings and enhanced tract variables.
\newblock In \emph{2024 32nd European Signal Processing Conference (EUSIPCO)}, pages 306--310. IEEE.

\bibitem[{Baevski et~al.(2020)Baevski, Zhou, Mohamed, and Auli}]{wav2vec2}
Alexei Baevski, Yuhao Zhou, Abdelrahman Mohamed, and Michael Auli. 2020.
\newblock wav2vec 2.0: A framework for self-supervised learning of speech representations.
\newblock \emph{Advances in neural information processing systems}, 33:12449--12460.

\bibitem[{Chen et~al.(2022)Chen, Wang, Chen, Wu, Liu, Chen, Li, Kanda, Yoshioka, Xiao et~al.}]{wavlm}
Sanyuan Chen, Chengyi Wang, Zhengyang Chen, Yu~Wu, Shujie Liu, Zhuo Chen, Jinyu Li, Naoyuki Kanda, Takuya Yoshioka, Xiong Xiao, et~al. 2022.
\newblock Wavlm: Large-scale self-supervised pre-training for full stack speech processing.
\newblock \emph{IEEE Journal of Selected Topics in Signal Processing}, 16(6):1505--1518.

\bibitem[{Chen et~al.(2021)Chen, Hung, Chuang, Sherman, Huang, Lu, and Tsao}]{EMA2S}
Yu-Wen Chen, Kuo-Hsuan Hung, Shang-Yi Chuang, Jonathan Sherman, Wen-Chin Huang, Xugang Lu, and Yu~Tsao. 2021.
\newblock Ema2s: An end-to-end multimodal articulatory-to-speech system.
\newblock In \emph{IEEE International Symposium on Circuits and Systems (ISCAS)}, pages 1--5.

\bibitem[{Cho et~al.(2024{\natexlab{a}})Cho, Mohamed, Black, and Anumanchipalli}]{universal-aai}
Cheol~Jun Cho, Abdelrahman Mohamed, Alan~W Black, and Gopala~K Anumanchipalli. 2024{\natexlab{a}}.
\newblock Self-supervised models of speech infer universal articulatory kinematics.
\newblock In \emph{ICASSP 2024-2024 IEEE International Conference on Acoustics, Speech and Signal Processing (ICASSP)}, pages 12061--12065. IEEE.

\bibitem[{Cho et~al.(2023)Cho, Wu, Mohamed, and Anumanchipalli}]{ssl-aai}
Cheol~Jun Cho, Peter Wu, Abdelrahman Mohamed, and Gopala~K Anumanchipalli. 2023.
\newblock Evidence of vocal tract articulation in self-supervised learning of speech.
\newblock In \emph{ICASSP 2023-2023 IEEE International Conference on Acoustics, Speech and Signal Processing (ICASSP)}, pages 1--5. IEEE.

\bibitem[{Cho et~al.(2024{\natexlab{b}})Cho, Wu, Prabhune, Agarwal, and Anumanchipalli}]{sparc}
Cheol~Jun Cho, Peter Wu, Tejas~S. Prabhune, Dhruv Agarwal, and Gopala~K. Anumanchipalli. 2024{\natexlab{b}}.
\newblock Coding speech through vocal tract kinematics.
\newblock \emph{IEEE Journal of Selected Topics in Signal Processing}, 18(8):1427--1440.

\bibitem[{Choi et~al.(2021)Choi, Lee, Kim, Lee, Heo, and Lee}]{nansy}
Hyeong-Seok Choi, Juheon Lee, Wansoo Kim, Jie Lee, Hoon Heo, and Kyogu Lee. 2021.
\newblock Neural analysis and synthesis: Reconstructing speech from self-supervised representations.
\newblock \emph{Advances in Neural Information Processing Systems}, 34:16251--16265.

\bibitem[{Chou et~al.(2019)Chou, Yeh, and Lee}]{instance_norm}
Ju-chieh Chou, Cheng-chieh Yeh, and Hung-yi Lee. 2019.
\newblock One-shot voice conversion by separating speaker and content representations with instance normalization.
\newblock \emph{arXiv preprint arXiv:1904.05742}.

\bibitem[{Du et~al.(2024{\natexlab{a}})Du, Chen, Zhang, Hu, Lu, Yang, Hu, Zheng, Gu, Ma et~al.}]{cosyvoice}
Zhihao Du, Qian Chen, Shiliang Zhang, Kai Hu, Heng Lu, Yexin Yang, Hangrui Hu, Siqi Zheng, Yue Gu, Ziyang Ma, et~al. 2024{\natexlab{a}}.
\newblock Cosyvoice: A scalable multilingual zero-shot text-to-speech synthesizer based on supervised semantic tokens.
\newblock \emph{arXiv preprint arXiv:2407.05407}.

\bibitem[{Du et~al.(2024{\natexlab{b}})Du, Wang, Chen, Shi, Lv, Zhao, Gao, Yang, Gao, Wang et~al.}]{cosyvoice2}
Zhihao Du, Yuxuan Wang, Qian Chen, Xian Shi, Xiang Lv, Tianyu Zhao, Zhifu Gao, Yexin Yang, Changfeng Gao, Hui Wang, et~al. 2024{\natexlab{b}}.
\newblock Cosyvoice 2: Scalable streaming speech synthesis with large language models.
\newblock \emph{arXiv preprint arXiv:2412.10117}.

\bibitem[{Gao et~al.(2024)Gao, Birkholz, and Li}]{vt_lab}
Yingming Gao, Peter Birkholz, and Ya~Li. 2024.
\newblock Articulatory copy synthesis based on the speech synthesizer vocaltractlab and convolutional recurrent neural networks.
\newblock \emph{IEEE/ACM Transactions on Audio, Speech, and Language Processing}.

\bibitem[{Hsu et~al.(2021)Hsu, Bolte, Tsai, Lakhotia, Salakhutdinov, and Mohamed}]{hubert}
Wei-Ning Hsu, Benjamin Bolte, Yao-Hung~Hubert Tsai, Kushal Lakhotia, Ruslan Salakhutdinov, and Abdelrahman Mohamed. 2021.
\newblock Hubert: Self-supervised speech representation learning by masked prediction of hidden units.
\newblock \emph{IEEE/ACM transactions on audio, speech, and language processing}, 29:3451--3460.

\bibitem[{Ju et~al.(2024)Ju, Wang, Shen, Tan, Xin, Yang, Liu, Leng, Song, Tang et~al.}]{naturalspeech3}
Zeqian Ju, Yuancheng Wang, Kai Shen, Xu~Tan, Detai Xin, Dongchao Yang, Yanqing Liu, Yichong Leng, Kaitao Song, Siliang Tang, et~al. 2024.
\newblock Naturalspeech 3: Zero-shot speech synthesis with factorized codec and diffusion models.
\newblock \emph{arXiv preprint arXiv:2403.03100}.

\bibitem[{Kameoka et~al.(2018)Kameoka, Kaneko, Tanaka, and Hojo}]{stargan-vc}
Hirokazu Kameoka, Takuhiro Kaneko, Kou Tanaka, and Nobukatsu Hojo. 2018.
\newblock Stargan-vc: Non-parallel many-to-many voice conversion using star generative adversarial networks.
\newblock In \emph{2018 IEEE Spoken Language Technology Workshop (SLT)}, pages 266--273. IEEE.

\bibitem[{Kaneko and Kameoka(2018)}]{cyclegan-vc}
Takuhiro Kaneko and Hirokazu Kameoka. 2018.
\newblock Cyclegan-vc: Non-parallel voice conversion using cycle-consistent adversarial networks.
\newblock In \emph{2018 26th European Signal Processing Conference (EUSIPCO)}, pages 2100--2104. IEEE.

\bibitem[{Kaneko et~al.(2019{\natexlab{a}})Kaneko, Kameoka, Tanaka, and Hojo}]{cyclegan-vc2}
Takuhiro Kaneko, Hirokazu Kameoka, Kou Tanaka, and Nobukatsu Hojo. 2019{\natexlab{a}}.
\newblock Cyclegan-vc2: Improved cyclegan-based non-parallel voice conversion.
\newblock In \emph{ICASSP 2019-2019 IEEE International Conference on Acoustics, Speech and Signal Processing (ICASSP)}, pages 6820--6824. IEEE.

\bibitem[{Kaneko et~al.(2019{\natexlab{b}})Kaneko, Kameoka, Tanaka, and Hojo}]{stargan-vc2}
Takuhiro Kaneko, Hirokazu Kameoka, Kou Tanaka, and Nobukatsu Hojo. 2019{\natexlab{b}}.
\newblock Stargan-vc2: Rethinking conditional methods for stargan-based voice conversion.
\newblock \emph{arXiv preprint arXiv:1907.12279}.

\bibitem[{Kashkin et~al.(2022)Kashkin, Karpukhin, and Shishkin}]{hifivc}
Anton Kashkin, Ivan Karpukhin, and Svyatoslav Shishkin. 2022.
\newblock Hifi-vc: High quality asr-based voice conversion.
\newblock \emph{arXiv preprint arXiv:2203.16937}.

\bibitem[{Kim et~al.(2018)Kim, Salamon, Li, and Bello}]{crepe}
Jong~Wook Kim, Justin Salamon, Peter Li, and Juan~Pablo Bello. 2018.
\newblock Crepe: A convolutional representation for pitch estimation.
\newblock In \emph{2018 IEEE international conference on acoustics, speech and signal processing (ICASSP)}, pages 161--165. IEEE.

\bibitem[{Kim et~al.(2023)Kim, Piao, Lee, and Kang}]{style}
Miseul Kim, Zhenyu Piao, Jihyun Lee, and Hong-Goo Kang. 2023.
\newblock Style modeling for multi-speaker articulation-to-speech.
\newblock In \emph{ICASSP}, pages 1--5.

\bibitem[{Koizumi et~al.(2023)Koizumi, Zen, Karita, Ding, Yatabe, Morioka, Bacchiani, Zhang, Han, and Bapna}]{libritts_r}
Yuma Koizumi, Heiga Zen, Shigeki Karita, Yifan Ding, Kohei Yatabe, Nobuyuki Morioka, Michiel Bacchiani, Yu~Zhang, Wei Han, and Ankur Bapna. 2023.
\newblock Libritts-r: A restored multi-speaker text-to-speech corpus.
\newblock \emph{arXiv preprint arXiv:2305.18802}.

\bibitem[{Kong et~al.(2020)Kong, Kim, and Bae}]{hifigan}
Jungil Kong, Jaehyeon Kim, and Jaekyoung Bae. 2020.
\newblock Hifi-gan: Generative adversarial networks for efficient and high fidelity speech synthesis.
\newblock \emph{Advances in neural information processing systems}, 33:17022--17033.

\bibitem[{Li et~al.(2023)Li, Tu, and Xiao}]{freevc}
Jingyi Li, Weiping Tu, and Li~Xiao. 2023.
\newblock Freevc: Towards high-quality text-free one-shot voice conversion.
\newblock In \emph{ICASSP 2023-2023 IEEE International Conference on Acoustics, Speech and Signal Processing (ICASSP)}, pages 1--5. IEEE.

\bibitem[{Lian et~al.(2022)Lian, Zhang, and Yu}]{d-dsvae}
Jiachen Lian, Chunlei Zhang, and Dong Yu. 2022.
\newblock Robust disentangled variational speech representation learning for zero-shot voice conversion.
\newblock In \emph{ICASSP 2022-2022 IEEE International Conference on Acoustics, Speech and Signal Processing (ICASSP)}, pages 6572--6576. IEEE.

\bibitem[{Liu et~al.(2024)Liu, Yu, Lin, Wu, Cho, and Anumanchipalli}]{slt2024}
Yisi Liu, Bohan Yu, Drake Lin, Peter Wu, Cheol~Jun Cho, and Gopala~Krishna Anumanchipalli. 2024.
\newblock Fast, high-quality and parameter-efficient articulatory synthesis using differentiable dsp.
\newblock In \emph{2024 IEEE Spoken Language Technology Workshop (SLT)}, pages 711--718. IEEE.

\bibitem[{Park et~al.(2020)Park, Kim, and Joe}]{cotatron}
Seung-won Park, Doo-young Kim, and Myun-chul Joe. 2020.
\newblock Cotatron: Transcription-guided speech encoder for any-to-many voice conversion without parallel data.
\newblock \emph{arXiv preprint arXiv:2005.03295}.

\bibitem[{Perez et~al.(2018)Perez, Strub, De~Vries, Dumoulin, and Courville}]{film}
Ethan Perez, Florian Strub, Harm De~Vries, Vincent Dumoulin, and Aaron Courville. 2018.
\newblock Film: Visual reasoning with a general conditioning layer.
\newblock In \emph{Proceedings of the AAAI conference on artificial intelligence}, volume~32.

\bibitem[{Qian et~al.(2020)Qian, Zhang, Chang, Hasegawa-Johnson, and Cox}]{triple_bottleneck}
Kaizhi Qian, Yang Zhang, Shiyu Chang, Mark Hasegawa-Johnson, and David Cox. 2020.
\newblock Unsupervised speech decomposition via triple information bottleneck.
\newblock In \emph{International Conference on Machine Learning}, pages 7836--7846. PMLR.

\bibitem[{Qian et~al.(2019)Qian, Zhang, Chang, Yang, and Hasegawa-Johnson}]{autovc}
Kaizhi Qian, Yang Zhang, Shiyu Chang, Xuesong Yang, and Mark Hasegawa-Johnson. 2019.
\newblock Autovc: Zero-shot voice style transfer with only autoencoder loss.
\newblock In \emph{International Conference on Machine Learning}, pages 5210--5219. PMLR.

\bibitem[{Qian et~al.(2022)Qian, Zhang, Gao, Ni, Lai, Cox, Hasegawa-Johnson, and Chang}]{contentvec}
Kaizhi Qian, Yang Zhang, Heting Gao, Junrui Ni, Cheng-I Lai, David Cox, Mark Hasegawa-Johnson, and Shiyu Chang. 2022.
\newblock Contentvec: An improved self-supervised speech representation by disentangling speakers.
\newblock In \emph{International conference on machine learning}, pages 18003--18017. PMLR.

\bibitem[{Siriwardena and Espy-Wilson(2023)}]{source_aai}
Yashish~M Siriwardena and Carol Espy-Wilson. 2023.
\newblock The secret source: Incorporating source features to improve acoustic-to-articulatory speech inversion.
\newblock In \emph{ICASSP 2023-2023 IEEE International Conference on Acoustics, Speech and Signal Processing (ICASSP)}, pages 1--5. IEEE.

\bibitem[{Sun et~al.(2016)Sun, Li, Wang, Kang, and Meng}]{ppg-vc}
Lifa Sun, Kun Li, Hao Wang, Shiyin Kang, and Helen Meng. 2016.
\newblock Phonetic posteriorgrams for many-to-one voice conversion without parallel data training.
\newblock In \emph{2016 IEEE International Conference on Multimedia and Expo (ICME)}, pages 1--6. IEEE.

\bibitem[{Van~Niekerk et~al.(2022)Van~Niekerk, Carbonneau, Za{\"\i}di, Baas, Seut{\'e}, and Kamper}]{softvc}
Benjamin Van~Niekerk, Marc-Andr{\'e} Carbonneau, Julian Za{\"\i}di, Matthew Baas, Hugo Seut{\'e}, and Herman Kamper. 2022.
\newblock A comparison of discrete and soft speech units for improved voice conversion.
\newblock In \emph{ICASSP 2022-2022 IEEE International Conference on Acoustics, Speech and Signal Processing (ICASSP)}, pages 6562--6566. IEEE.

\bibitem[{Wei et~al.(2023)Wei, Cao, Dan, and Chen}]{rmvpe}
Haojie Wei, Xueke Cao, Tangpeng Dan, and Yueguo Chen. 2023.
\newblock Rmvpe: A robust model for vocal pitch estimation in polyphonic music.
\newblock \emph{arXiv preprint arXiv:2306.15412}.

\bibitem[{Wu et~al.(2023)Wu, Chen, Cho, Watanabe, Goldstein, Black, and Anumanchipalli}]{peter_aai}
Peter Wu, Li-Wei Chen, Cheol~Jun Cho, Shinji Watanabe, Louis Goldstein, Alan~W Black, and Gopala~K. Anumanchipalli. 2023.
\newblock Speaker-independent acoustic-to-articulatory speech inversion.
\newblock In \emph{ICASSP 2023 - 2023 IEEE International Conference on Acoustics, Speech and Signal Processing (ICASSP)}, pages 1--5.

\bibitem[{Wu et~al.(2021)Wu, Liang, Shi, Salakhutdinov, Watanabe, and Morency}]{peter-vc}
Peter Wu, Paul~Pu Liang, Jiatong Shi, Ruslan Salakhutdinov, Shinji Watanabe, and Louis-Philippe Morency. 2021.
\newblock Understanding the tradeoffs in client-side privacy for downstream speech tasks.
\newblock In \emph{2021 Asia-Pacific Signal and Information Processing Association Annual Summit and Conference (APSIPA ASC)}, pages 841--848. IEEE.

\bibitem[{Wu et~al.(2022)Wu, Watanabe, Goldstein, Black, and Anumanchipalli}]{Peter-ATS}
Peter Wu, Shinji Watanabe, Louis Goldstein, Alan~W Black, and Gopala~Krishna Anumanchipalli. 2022.
\newblock Deep speech synthesis from articulatory representations.
\newblock In \emph{Interspeech}.

\bibitem[{Yamagishi et~al.(2019)Yamagishi, Veaux, and MacDonald}]{vctk}
Junichi Yamagishi, Christophe Veaux, and Kirsten MacDonald. 2019.
\newblock \href {https://doi.org/10.7488/ds/2645} {Cstr vctk corpus: English multi-speaker corpus for cstr voice cloning toolkit (version 0.92)}.
\newblock [sound].

\bibitem[{Yang et~al.(2024)Yang, Kartynnik, Li, Tang, Li, Sung, and Grundmann}]{streamvc}
Yang Yang, Yury Kartynnik, Yunpeng Li, Jiuqiang Tang, Xing Li, George Sung, and Matthias Grundmann. 2024.
\newblock \href {https://arxiv.org/abs/2401.03078} {Streamvc: Real-time low-latency voice conversion}.
\newblock \emph{Preprint}, arXiv:2401.03078.

\bibitem[{Zeghidour et~al.(2021)Zeghidour, Luebs, Omran, Skoglund, and Tagliasacchi}]{soundstream}
Neil Zeghidour, Alejandro Luebs, Ahmed Omran, Jan Skoglund, and Marco Tagliasacchi. 2021.
\newblock Soundstream: An end-to-end neural audio codec.
\newblock \emph{IEEE/ACM Transactions on Audio, Speech, and Language Processing}, 30:495--507.

\bibitem[{Zen et~al.(2019)Zen, Dang, Clark, Zhang, Weiss, Jia, Chen, and Wu}]{libritts}
Heiga Zen, Viet Dang, Rob Clark, Yu~Zhang, Ron~J Weiss, Ye~Jia, Zhifeng Chen, and Yonghui Wu. 2019.
\newblock Libritts: A corpus derived from librispeech for text-to-speech.
\newblock \emph{arXiv preprint arXiv:1904.02882}.

\end{thebibliography}




\end{document}